\documentclass[a4paper,11pt]{article}
\usepackage{jheppub} 
\usepackage{lineno}
\usepackage{amssymb} 
\usepackage[section]{placeins}
\usepackage{subcaption}

\title{\boldmath Quark spin-orbit correlations in spin-0 and spin-1 mesons using the light-front quark model}

\author{Ritwik Acharyya$^{1}$, Satyajit Puhan$^{1}$ and Harleen Dahiya$^{1}$}
\affiliation{$^{1}$Department of Physics, Dr. B. R. Ambedkar National Institute of Technology Jalandhar-144008, India}
\emailAdd{ritwikacharyya2001@gmail.com}
\emailAdd{puhansatyajit@gmail.com}
\emailAdd{dahiyah@nitj.ac.in}

\abstract{We have investigated the spin-orbital angular momentum correlations for the active quark inside the light and heavy mesons for both the spin-0 and spin-1 cases. These correlations can be derived from the  generalised transverse momentum dependent distributions (GTMDs) as well as the generalised parton distributions (GPDs). We employ the overlap representation of light-front wave functions in the light-front quark model (LFQM) to calculate our analytical results. The dependence of spin-orbit correlations (SOCs) on the longitudinal momentum fraction $x$ as well as the transverse momentum dependence $\mathbf{k}_{\perp}$ has been graphically presented. Even though the SOCs have already been studied for the spin-0 pions and kaons in other approaches, no calculations for the other light and heavy spin-0 mesons have been reported in literature. Further, the correlations for any of the light and heavy spin-1 mesons have been studied for the first time in the present work.}


\begin{document}
\maketitle

\flushbottom

\section{INTRODUCTION} {\label{sec:intro}}

Quantum Chromodynamics (QCD) \cite{Marciano:1977su,Arifi:2022pal, Altarelli:1981ax} describes the production of hadrons by incorporating strong interactions among quarks, anti-quarks, and gluons. One of the most significant problems and an  avenue for improving our comprehension of QCD and confinement is to decipher the multidimensional structure of hadron thus providing insight into several non-perturbative aspects of QCD.  Partons inside the hadrons possess a specific momentum and position and their state can be described by Wigner distributions which are six-dimensional phase-space distributions \cite{Bacchetta:2008af, Lorce:2011kd}. Wigner distributions are the quantum-mechanical constructions that are closest to a classical probability density in phase-space. The probability density of discovering a parton (gluon or quark) carrying the parent hadron's light-front (LF) longitudinal momentum fraction $x$ is described by the parton distribution function (PDF) \cite{Han:2022tlh,Placakyte:2011az,martin1998parton,Bacchetta:2008af,Gluck:1998xa,Gluck:1994uf,Mulders:1995dh}. For the description of observables that are also sensitive to the transverse kinematics of a parton, the concept of PDFs has been extended to transverse momentum dependent parton distributions (TMDs) \cite{Puhan:2023hio,Angeles-Martinez:2015sea, Boer:1997nt,Boer:1999mm, Ralston:1979ys,Kotzinian:1994dv,Radici:2014bfa,Sivers:1989cc} and generalized parton distributions (GPDs) \cite{Garcon:2002jb,Belitsky:2005qn,Diehl:2015uka,Chakrabarti:2005zm,Brodsky:2006ku,Rajan:2016tlg,Hagler:2003jw,Ji:2004gf} to include the information of transverse momentum and transverse coordinate distributions respectively. After a few phase-space reductions, Wigner distributions reduce to TMDs and GPDs. Both TMDs and GPDs present a three-dimensional (3-D) visualization of the hadron. The Wigner distributions integrated over the transverse momenta reduce to the GPDs at zero skewness $(\zeta = 0)$ \cite{Chakrabarti:2016yuw}. On the other hand, integrating it over the transverse impact parameter, with zero momentum transfer, they reduce to the TMDs. TMDs can be measured in certain reactions like semi-inclusive deep inelastic scattering (SIDIS) \cite{Bacchetta:2017gcc,brodsky1998quantum,Dirac:1949cp}, Drell-Yan (DY) processes \cite{Collins:2002kn,Ralston:1979ys,Zhou:2009jm,Tangerman:1994eh,Donohue:1980tn}  and $Z^{0}/ W^{\pm}$ production \cite{Catani:2015vma,Scimemi:2017etj,Collins:1984kg}. GPDs are extracted from the QCD description of hard exclusive reactions like deeply virtual Compton scattering (DVCS) \cite{muller1994wave,belitsky2005unraveling,Boffi:2007yc,Goeke:2001tz,Radyushkin:1996nd} and deeply virtual meson production (DVMP) \cite{Mankiewicz:1997uy,Frankfurt:1999fp}. Analysing the GPDs can reveal details about the quarks' spatial distributions.  Further, if we integrate the Wigner distribution over the LF energy of the parton, it can be interpreted as a Fourier transform of corresponding generalized transverse momentum dependeny distributions (GTMDs) \cite{Lorce:2013pza,Meissner:2009ww} which are functions of the LF three momentum of the parton as well as the momentum transfer to the hadron \cite{Chakrabarti:2016yuw}. 

The spin-orbit correlations (SOCs) between a hadron and a quark can be explained on the basis of the phase-space average of Wigner distributions. Comprehending spin composition of hadrons has been a fascinating subject of research lately \cite{Ji:1996ek,Leader:2013jra,Kaur:2020vkq, Jaffe:1989jz} and understanding the multidimensional structure makes it possible to analyse characteristics such as SOCs, spin-spin correlations, quark-gluon correlations and other such interactions. It would be interesting to explore the connection of the GPDs and GTMDs of spin-0 and spin-1 mesons with the spin-orbital angular momentum correlations for them and proceed to calculate the analytical results for them. In particular, the correlations between the hadron spin and the orbital motion of partons inside the hadron can bring much bigger insight into the spin structure of the hadrons \cite{Tan:2021osk}. Given that a parton's orbital angular momentum (OAM) and spin contributions have intrinsic negative parity, the only non-vanishing single-parton ($a = q, G$) correlations allowed by parity invariance are $\mathbf{S}_{a} \cdot \mathbf{S}_{N}$, $\mathbf{L}_{a}$$\cdot$$\mathbf{S}_{N}$, and $\mathbf{S}_{a}$ $\cdot$ $\mathbf{L}_{a}$. Here, $\mathbf{L}_{q,G}$ represents the quark (or gluon) OAM, $\mathbf{S}_{N}$ is the spin of the hadron and $\mathbf{S}_{q,G}$ portrays the spin of the constituent quark (or gluon) \cite{Tan:2021osk}. The initial two types of correlation are commonly referred to as OAM and spin contributions of parton $a$ to the spin of the hadron whereas the third and final type is the parton's SOC. We use the difference between the right-handed and left-handed contributions of the quark longitudinal OAM to describe the quark longitudinal SOC, which is expressed by $C_{z}^{q}$. 

Dirac, in 1949 \cite{Harindranath:1996hq}, recognized that one may set up a dynamical theory in which the dynamical variables refer to the physical conditions on a front $x^{+} = 0$. The resulting dynamics is called the LF dynamics, which Dirac referred to as \emph{front-form} for brevity. The LF dynamics \cite{Terentev:1976jk,Arifi:2022qnd,brodsky1998quantum,Lepage:1980fj} is a beneficial model framework which helps us study the internal structure of hadrons \cite{Brodsky:1997de,Harindranath:1996hq} and has direct applications in the Minkowski space \cite{Jacobson:2012ei}. LF quantization provides a framework to describe the perturbative and non-perturbative regimes of QCD. LF dynamics can be realized by a number of different models and for this work we have adopted the LF quark model (LFQM) \cite{Jaus:1989au, Jaus:1991cy, Coester:2005cv}. LFQM is based on the algebra of generators of the Lorentz group in the LF dynamics \cite{Belyaev:1997iu}. The component quark and anti-quark in a bound state must be on-mass shell in conventional LFQM \cite{Arifi:2023jfe,Arifi:2022pal,deMelo:2003uk,deMelo:1997hh}. The spin-orbit wave function is derived from the conventional time independent spin-orbit wave function supplied by the quantum numbers $J^{PC}$ \cite{Choi:2021mni} using the well known Melosh transformation, which are independent of interactions \cite{Melosh:1974cu}. LFQM is primarily concerned with the valence quarks of hadrons, which are among the primary elements responsible for the overall composition and properties of hadrons. With accurate parameter choices, the model describes several hadron characteristics, including form factors for $Q^{2} = 1 \hspace{1mm} GeV^{2}$, thus establishing a phenomenological link between hadron properties and the wave function of the quark constituents which has been successful in many instances.

Quark SOCs have been studied earlier for the spin-$\frac{1}{2}$ hadrons \cite{Lorce:2014mxa} and also investigated for the case of spin-0 hadrons like pions and kaons \cite{Kaur:2019jow,Tan:2021osk}. However, no work has been reported for the remaining members of the spin-0 mesons. Further, the spin-1 mesons (light and heavy) have also remained unexplored in this regard. We may obtain special insights on the orbital motion of quarks and their intrinsic longitudinal spin inside spin-0 and spin-1 mesons owing to the quark SOC \cite{Tan:2021osk}. In light of the successes of the LFQM and the importance of the quark SOCs, it becomes essential to extend this work across all the members in both the spin-0 and spin-1 meson spectrum. To make the application of this work broader, we have included the light as well as heavy mesons for both the spin-0 and spin-1 cases.  We have utilized the conventional definition of the leading-twist GTMD $G_{1,1}$ in our calculations and have solved the correlators for the leading-twist GPD case. To derive the outcome for the SOC, we have integrated the GTMD $G_{1,1}$ twice in terms of transverse momentum $\mathbf{k}_{\perp}$ and fraction of momentum transfer to the active quark $x$ involving the entire wave function of the LFQM. We have visualized the behavior of the spin-orbit correlator $C_{z}^{q}$ in our chosen model via two-dimensional (2-D) plots with respect to the longitudinal momentum fraction $x$ and transverse momenta of quark $\mathbf{k}_{\perp}$. We have also presented the model dependent results for the pion and kaon. Further, the physical implications of the SOC for both the spin-0 and spin-1 light and heavy mesons have been discussed. 

The paper is arranged as follows. In Section \ref{sec2}, we have presented the correlation between the quark spin and OAM inside the hadron. In Section \ref{sec3}, we have quantitatively discussed LFQM: the model employed to define our LF wave functions (LFWFs). In Section \ref{sec4}, we have shown how the spin-OAM correlations can be methodically derived from the GPDs and the GTMDs. These relations have been presented for both the light and heavy spin-0 and spin-1 mesons. Further, in Section \ref{sec5}, we have defined our model parameters and presented our model results for the SOC. Finally, we have summarized our results in Section \ref{sec6}.

\section{SPIN-ORBIT CORRELATION} {\label{sec2}}

The local gauge-invariant LF operators for the quark longitudinal spin and OAM have been of unique interest since they enter the Ji's decomposition of the total angular momentum operator in QCD \cite{Tan:2021osk, Ji:1996ek} which is given as
\begin{equation}
\hat{J}_{z} =  \hat{S}_{z}^{q} + \hat{L}_{z}^{q} + \hat{J}_{z}^{G}.
\end{equation}
Here, $\hat{L}_{z}^{q}$ refers to the gauge-invariant LF quark longitudinal OAM which can further be decomposed into right-handed and left-handed quark contributions as 
\begin{equation}
    \hat{L}_{z}^{q} = \int d^{3}x \frac{1}{2} \overline{\psi} \gamma^{+} (\mathrm{\mathbf{x}} \times i \overleftrightarrow{\mathbf{D}})_{z} \psi = \hat{L}_{z}^{qR} + \hat{L}_{z}^{qL},
\end{equation}
where the symmetric covariant derivative is defined by $\overleftrightarrow{\mathbf{D}} = \overleftarrow{\partial} - \overrightarrow{\partial} - 2ig\mathbf{A}$ \cite{Tan:2021osk}. $\psi_{R,L} = \frac{1}{2}(\mathbf{I} \pm \gamma_{5}) \psi$ and  $d^{3}x = dx^{-}d^{2}x_{\perp}$. The knowledge and understanding of quark SOCs give us a complete characterization of the hadron's internal structure. The local gauge-invariant correlation is described by
\begin{equation}
    \hat{C_{z}^{q}} =  \int d^{3}x \frac{1}{2} \overline{\psi} \gamma^{+} \gamma_{5} (\mathrm{\mathbf{x}} \times i \overleftrightarrow{\mathbf{D}})_{z} \psi = \hat{L}_{z}^{qR}-\hat{L}_{z}^{qL}.
\end{equation}
The quark OAM operator may be represented in terms of the gauge-invariant energy-momentum tensor as follows \cite{Lorce:2014mxa}
\begin{equation}
     \hat{L}_{z}^{q} = \int d^{3}x (x^{1} \hat{T_{q}^{+2}} - x^{2} \hat{T_{q}^{+1}}),
\end{equation}
where $\hat{T^{\mu\nu}}$ is the energy-momentum tensor operator given by
\begin{align}
    \hat{T^{\mu\nu}_{q}} &= \frac{1}{2} \overline{\psi} \gamma^{\mu} i \overleftrightarrow{\mathbf{D}}^{\nu} \psi \\
    &=  \hat{T_{qR}^{\mu\nu}} - \hat{T_{qL}^{\mu\nu}},
\end{align}
and $\hat{T^{\mu\nu}_{R,L}} = \frac{1}{2} \overline{\psi}_{R,L} \gamma^{\mu} i \overleftrightarrow{\mathbf{D}}^{\nu} \psi_{R,L}$. 
Further, the quark SOC operator is given as follows
\begin{equation} \label{eq:2.6}
     \hat{C_{z}^{q}} = \int d^{3}x (x^{1} \hat{T_{q5}^{+2}} - x^{2} \hat{T_{q5}^{+1}}),
\end{equation}
where $\hat{T^{\mu\nu}_{q5}}$ may be regarded as the parity-odd partner of $\hat{T^{\mu\nu}}$ and can be expressed as \cite{Lorce:2014mxa}
\begin{align}
\hat{T^{\mu\nu}_{q5}} &= \frac{1}{2} \overline{\psi} \gamma^{\mu} \gamma_{5} i \overleftrightarrow{\mathbf{D}}^{\nu} \psi \\
&=  \hat{T_{qR}^{\mu\nu}} + \hat{T_{qL}^{\mu\nu}}. 
\end{align}
The non-forward matrix components of $\hat{T^{\mu\nu}_{q5}}$, inserted between two meson states, may be parametrised as a sum of two form factors $\Tilde{\mathcal{C}_{q}}(t)$ and $\Tilde{\mathcal{F}_{q}}(t)$ 
\cite{Tanaka:2018wea,Polyakov:2018zvc,Krutov:2020ewr, Freese:2019bhb}, which can be expressed as
\begin{equation} \label{eq: 2.8}
    \langle k'\vert \hat{T^{\mu\nu}_{q5}} (0) \vert k \rangle = - \frac{P^{[\mu_{i\epsilon}\nu]+ \Delta P}}{2P^{+}} (\Tilde{\mathcal{C}_{q}}(t) - 2 \Tilde{\mathcal{F}_{q}}(t)) + i \epsilon^{\mu\nu\Delta P} \Tilde{\mathcal{F}_{q}}(t) + \mathcal{O}(\Delta^{2}).
\end{equation}
Substituting Eq. (\ref{eq: 2.8}) into the matrix elements of Eq. (\ref{eq:2.6}), within the symmetric LF frame ($\mathbf{P}_{\perp} = \mathbf{0}_{\perp}$), we get
\begin{equation}
    C_{z}^{q} = \frac{\langle p \vert \hat{C}_{z}^{q} \vert p \rangle}{\langle p \vert p \rangle} = \Tilde{\mathcal{C}}_{q}(0).
\end{equation}
Thus, we just need to compute the form factor $\Tilde{\mathcal{C}}_{q}(t)$ in order to determine the quark SOCs for the mesons.

\section{LIGHT-FRONT QUARK MODEL}{\label{sec3}}

In the LF technique, a sequence of LFWFs in the Fock-state basis are used to define the wave functions of the meson describing a composite state at a certain LF time \cite{Ma:2018ysi}. 
The meson eigenstate $\vert \mathbb{M}(P^{+}, \mathbf{P_{\perp}}, S_{z})\rangle$ can be expressed in terms of its component eigenstate $\vert n \rangle$ using the LF Fock state expansion and can be expressed  as \cite{Kaur:2019jow, Brodsky:1982nx,Chakrabarti:2016yuw}

\begin{eqnarray}{\label{3.2}}
    \vert \mathbb{M}(P^{+}, \mathbf{P_{\perp}}, S_{z})\rangle & = & \sum_{n, \lambda_{i}} \int \prod_{i=1}^{n} \frac{dx_{i} d^{2} \mathbf{k}_{\perp i}}{\sqrt{x_{i}} 16 \pi^{3}} 16 \pi^{3}  \delta \left ( 1 - \sum_{i=1}^{n} x^{i} \right ) \delta^{(2)} \left ( \sum_{i=1}^{n} \mathbf{k}_{\perp i} \right ) \nonumber \\
   && \vert n;x_{i}P^{+}, x_{i}\mathbf{P_{\perp}} +  \mathbf{k}_{\perp i}, \lambda_{i} \rangle \psi_{n} (x_{i}, \mathbf{k}_{\perp i}).
\end{eqnarray}
Here, we denote $P$ = $(P^{+}, P^{-}, P_{\perp})$ as the meson’s total momentum and $S_{z}$ as the longitudinal spin projection of the target. The LF momentum co-ordinates and relative momentum fractions of the mesonic components are denoted by $\mathbf{k}_{\perp i}$ and $x_{i} = k_{i}^{+}/P^{+}$ respectively. The quark's transverse and longitudinal momentum fractions are represented by the symbols $\mathbf{k}_{\perp}$ and $x$ respectively. In contrast, $1-x$ and $-\mathbf{k}_{\perp}$ respectively describe the same for the anti-quark spectator. $\lambda_{i}$ is the helicity and $m_{i}$ is the mass of the $\emph{i}^{th}$ constituent correspondingly. In Eq. (\ref{3.2}),  $x_{i}\mathbf{P_{\perp}} + \mathbf{k}_{\perp i} = \mathbf{p}_{\perp i}$ is the physical transverse momentum, $\psi_{n}$ is the state  describing the probability for calculating the respective meson's on-shell mass components \cite{Kaur:2019jow, Brodsky:2007hb}. To simplify our calculations, we have considered the minimal Fock-state description of meson in the form of a quark-antiquark pair and is expressed as  \cite{Hoyer:1988vq}

\begin{equation}
    \vert \mathbb{M}(P,S)\rangle = \sum_{\lambda_{1}, \lambda_{2}} \int \frac{dx d^{2}\mathbf{k}_{\perp}}{16 \pi^{3} \sqrt{x(1-x)}} \vert x, \mathbf{k}_{\perp}, \lambda_{1}, \lambda_{2} \rangle 
      \Psi_{\lambda_{1}, \lambda_{2}}^{\Lambda} (x, \mathbf{k}_{\perp}),
\end{equation}
where $\Psi_{\lambda_{1}, \lambda_{2}}^{\Lambda} (x, \mathbf{k}_{\perp})$ is the LFWF with different spin and helicity projections and $\lambda_{1 (2)}$ describes the helicity of the quark (anti-quark) in the meson. Also, $\Lambda = T$ refers to the transverse spin projections of the mesons. The momenta of the meson ($P$), constituent quark ($k_{1}$) and anti-quark ($k_{2}$) in LFQM are given as

\begin{eqnarray}
     P &= &\left ( P^{+}, \frac{M^{2}}{P^{+}}, \mathbf{0}_{\perp}  \right ), \\
     k_{1} &= &\left ( xP^{+}, \frac{\mathbf{k}_{\perp}^{2} + m^{2}_{q}}{xP^{+}}, \mathbf{k}_{\perp} \right ), \\
     k_{2} &= &\left ( (1-x)P^{+}, \frac{\mathbf{k}_{\perp}^{2} + m^{2}_{\overline{q}}}{(1-x)P^{+}}, - \mathbf{k}_{\perp} \right ).
\end{eqnarray}
Here, $m_{q}$($m_{\overline{q}}$) refers to the boost invariant mass of the quark (anti-quark) and $M$ refers to the mass of the meson which is given by
\begin{equation}
    M^2 = {\frac{\mathbf{k}_{\perp}^{2} + m_{q}^2}{x}} + {\frac{\mathbf{k}_{\perp}^{2} + m_{\overline{q}}^2}{1-x}}.
\end{equation}
The LF meson wave function is expressed as \cite{Yu:2007hp,Qian:2008px}

\begin{equation} \label{eq3.7}
\Psi_{\lambda_{1}, \lambda_{2}}^{\Lambda} (x, \mathbf{k}_{\perp}) = \phi (x,\mathbf{k}_{\perp}) X_{\lambda_{1}, \lambda_{2}}^{\Lambda} (x, \mathbf{k}_{\perp}).
\end{equation} 
Here, $X_{\lambda_{1}, \lambda_{2}}^{\Lambda} (x, \mathbf{k}_{\perp})$ represents the spin wave function and $\phi (x,\mathbf{k}_{\perp})$ is the momentum-space wave function of the meson, respectively.
Let us begin our discussions with spin-0 meson. The spin wave functions is derived through the Melosh-Wigner rotation \cite{Diehl:2003ny, Liu:2015eqa, Ma:1992sj} and for spin-0 mesons, they are expressed as \cite{Puhan:2023ekt}

\begin{equation}
    X^{\mathcal{SP}0} (x, \mathbf{k}_{\perp}) = \sum_{\lambda_{1}, \lambda_{2}} K_{0} (x, \mathbf{k}_{\perp}, \lambda_{1}, \lambda_{2}) X_{1}^{\lambda_{1}} X_{2}^{\lambda_{2}},
\end{equation}
where $\mathcal{SP}0$ stands for spin-0 meson. $K_{0} (x, \mathbf{k}_{\perp}, \lambda_{1}, \lambda_{2})$ is the coefficient of the spin wave function. The different helicity combinations are described as 
\begin{eqnarray}
    K_{0} (x, \mathbf{k}_{\perp}, \uparrow, \downarrow) &=& [(xM + m_{q})((1-x)M + m_{\bar{q}}) - k_{\perp}^{2}] / \sqrt{2} w_{1} w_{2}, \\
    K_{0} (x, \mathbf{k}_{\perp}, \downarrow, \uparrow) &=& -[(xM + m_{q})((1-x)M + m_{\bar{q}}) - k_{\perp}^{2}] / \sqrt{2} w_{1} w_{2}, \\
    K_{0} (x, \mathbf{k}_{\perp}, \uparrow, \uparrow) &=& [(xM + m_{q})k_{2}^{L} - ((1-x)M + m_{\bar{q}}) k_{1}^{L}] / \sqrt{2} w_{1} w_{2}, \\
    K_{0} (x, \mathbf{k}_{\perp}, \downarrow, \downarrow) &=& [(xM + m_{q})k_{2}^{R} - ((1-x)M + m_{\bar{q}}) k_{1}^{R}] / \sqrt{2} w_{1} w_{2}.
\end{eqnarray}
Here, the subscripts $\uparrow$ and $\downarrow$ denote the transverse polarizations of the quark along the directions $\hat{e}_{x}$ and  $-\hat{e}_{x}$, respectively and 
\begin{eqnarray}
    w_{1} &=& [(xM + m_{q})^{2} + \mathbf{k}_{\perp}^2]^{\frac{1}{2}},\\
   w_{2} &=& [((1-x)M + m_{\overline{q}})^{2} + \mathbf{k}_{\perp}^2]^{\frac{1}{2}}.
\end{eqnarray}

For the spin-1 mesons, the spin wave functions occurring in Eq. (\ref{eq3.7}) take the form for $\Lambda$ = T(+) as \cite{Kaur:2020emh}
\begin{eqnarray}
    X_{+,+}^{T(+)} (x, \mathbf{k}_{\perp}) &=& \frac{m_{q} (M + 2m) + \mathbf{k}_{\perp}^{2}}{(M + m_{q}) \sqrt{ m_{q}^{2} + \mathbf{k}_{\perp}^{2}}}, \\
    X_{-,+}^{T(+)} (x, \mathbf{k}_{\perp}) &=& -  \frac{ k_{R}((1-x) M + m_{q}) }{(M + m_{q}) \sqrt{ m_{q}^{2} + \mathbf{k}_{\perp}^{2}}},\\
    X_{+,-}^{T(+)} (x, \mathbf{k}_{\perp}) &=& \frac{ k_{R}(x M + m_{q})}{(M + m_{q}) \sqrt{ m_{q}^{2} + \mathbf{k}_{\perp}^{2}}}, \\
    X_{-,-}^{T(+)} (x, \mathbf{k}_{\perp}) &=& - \frac{k_{R}^{2}}{(M + m_{q}) \sqrt{ m_{q}^{2} + \mathbf{k}_{\perp}^{2}}}.
\end{eqnarray}
Similarly, the spin wave functions occurring in Eq. (\ref{eq3.7}) takes the form for $\Lambda$ = T(-) as

\begin{eqnarray}
    X_{+,+}^{T(-)} (x, \mathbf{k}_{\perp}) &=& - \frac{k_{L}^{2}}{(m_{q} + M) \sqrt{ m_{q}^{2} + \mathbf{k}_{\perp}^{2}}}, \\
    X_{-,+}^{T(-)} (x, \mathbf{k}_{\perp}) &=& - \frac{(xM + m_{q}) k_{L}}{(M + m_{q}) \sqrt{ m_{q}^{2} + \mathbf{k}_{\perp}^{2}}}, \\
    X_{+,-}^{T(-)} (x, \mathbf{k}_{\perp}) &=&  \frac{((1-x) M + m_{q}) k_{L}}{(M + m_{q}) \sqrt{ m_{q}^{2} + \mathbf{k}_{\perp}^{2}}}, \\
    X_{-,-}^{T(-)} (x, \mathbf{k}_{\perp}) &=& \frac{m_{q} (M + 2m) + \mathbf{k}_{\perp}^{2}}{(M + m_{q}) \sqrt{ m_{q}^{2} + \mathbf{k}_{\perp}^{2}}},
\end{eqnarray}
where
\begin{equation*}
    k_{R(L)} = k_{x} \pm ik_{y}.
\end{equation*}
The momentum-space wave function can be described using the Brodsky-Huang-Lepage method as \cite{Xiao:2002iv} 
\begin{equation}{\label{3.13}}
\phi (x,\mathbf{k}_{\perp}) = A \exp{\left [- \frac{\frac{m_{q}^{2} + \mathbf{k}_{\perp}^{2}}{x} + \frac{m_{\overline{q}}^{2} + \mathbf{k}_{\perp}^{2}}{1-x}}{8\beta^{2}} - \frac{(m_{q}^{2} - m_{\overline{q}}^{2})^{2}}{8\beta^{2}(\frac{m_{q}^{2} + \mathbf{k}_{\perp}^{2}}{x} + \frac{m_{\overline{q}}^{2} + \mathbf{k}_{\perp}^{2}}{1-x})} \right]}.
\end{equation} 
Here, $\beta$  refers to the harmonic oscillator (HO) scale parameter and $A$ to the normalization constant. For $m_{q} = m_{\overline{q}} = m$, we have pion-like mesons, and for $m_{q} \neq m_{\overline{q}}$, we have kaon-like mesons.

\section{CONNECTING SOC TO GPDs and GTMDs}{\label{sec4}}

The energy-momentum tensor operator $\hat{T^{\mu\nu}}$ has no fundamental probe that can couple to its parity-odd partner $\hat{T^{\mu\nu}_{q5}}$ in high energy physics. However, by connecting the respective form factors (FF) to the exact moments of the GTMDs or GPDs we can obtain a representation of $\hat{T^{\mu\nu}_{q5}}$ \cite{Tan:2021osk}. The relationship between the FFs  may be derived employing the following QCD relation
\begin{equation} \label{momtensor}
\overline{\phi} \gamma^{[\mu} \gamma_{5} i \overleftrightarrow{\mathbf{D}}^{\nu]} \phi = 2m\overline{\phi} i \sigma^{\mu\nu} \gamma_{5} \phi - \epsilon^{\mu\nu\alpha\beta} \partial_{\alpha} (\overline{\psi} \gamma_{\beta} \phi).
\end{equation} 
When we focus on the matrix's off-diagonal components in the equation given above, the left-hand side represents the SOC \cite{Tan:2021osk} whereas the right-hand side parameterises the vector and tensor local correlators as \cite{Lorce:2014mxa}
\begin{align}
     \langle p'\vert \overline{\phi} \gamma^{\mu} \phi \vert p \rangle &= \Gamma^{\mu}_{qV},
\end{align} 
\begin{align}
\langle p'\vert \overline{\phi} i \sigma^{\mu\nu} \gamma_{5} \phi \vert p \rangle &= \Gamma^{\mu\nu}_{qT},
\end{align} 
where 
\begin{eqnarray}
	\Gamma^{\mu\nu}_{qT} &=& \frac{2i \epsilon^{\mu\nu\alpha\beta} \Delta_{\alpha} P_{\beta}}{M} \int H_{1}^{q} (x, \zeta, t) dx,\\
    \Gamma^{\mu}_{qV} &=& 2 P^{\mu} \int F_{1}^{q} (x, \zeta, t) dx.
\end{eqnarray}
Here, $\zeta = - \Delta^{+}/2P^{+}$ is the skewness variable with $\Delta$ being the momentum transfer. The functions $H_{1}^{q} (x, \zeta, t)=H_1$(for simplicity) and $F_{1}^{q} (x, \zeta, t)=F_1$(for simplicity) are defined as GPDs \cite{Burkardt:2007xm,Metz:2009zz} of the meson. $H_{1}^{q} (x, \zeta, t)$ represents the axial-vector LF quark correlator and $F_{1}^{q} (x, \zeta, t)$ represents the tensor LF quark correlator.  They are given as
\begin{align}
       \frac{1}{2} \int\frac{dy^{-}}{2\pi} e^{ixP^{+}z^{-}} \langle p' \vert \overline{\phi} \left (-\frac{y^{-}}{2} \right ) i\sigma^{j+} \gamma_{5} \phi \left (\frac{y^{-}}{2} \right) \vert p \rangle &= -\frac{i \epsilon^{ij}_{\perp} \Delta^{i}_{\perp}}{M} H_{1}^{q} (x, \zeta, t), \\
       \frac{1}{2} \int\frac{dy^{-}}{2\pi} e^{ixP^{+}y^{-}} \langle p' \vert \overline{\phi} \left (-\frac{y^{-}}{2} \right ) \gamma^{+} \phi \left (\frac{y^{-}}{2} \right) \vert p \rangle &= F_{1}^{q} (x, \zeta, t). 
    \end{align}
Hence, the SOC may be ascertained through the combinations of the moments of $F_{1}^{q} (x, \zeta, t)$ and $H_{1}^{q} (x, \zeta, t)$
\begin{equation}
\Tilde{C}_{q}(t) = \int dx \left (\frac{m_{q}}{M} H_{1}^{q} (x, \zeta, t) - \frac{1}{2} F_{1}^{q} (x, \zeta, t) \right).
\end{equation} 
Therefore, the expectation value of the SOC can be expressed analytically in the form of GPDs as 
\begin{equation}
{C}^{q}_{z} = \int dx \left (\frac{m_{q}}{M} H_{1}^{q} (x, 0, 0) - \frac{1}{2} F_{1}^{q} (x, 0, 0) \right).
\end{equation}

Further, the SOC can also be expressed in the form of GTMDs \cite{Chakrabarti:2016yuw, Lorce:2011kd, Kaur:2019jow}. We can exhibit $C_{z}^{q}$ in the form of one of the leading twist-2 GTMDs  $G_{1,1}\left(x, \zeta, \boldsymbol{k}_{\perp}^{2}, \boldsymbol{k}_{\perp} \cdot \boldsymbol{\Delta}_{\perp}, \boldsymbol{\Delta}_{\perp}^{2}\right)=G_{1,1}$(for simplicity) that are related to unpolarized meson states. For the present work, we consider the case of zero skewness i.e., $\zeta = 0$. We have
\begin{equation} \label{eq:4.10}
    {C}^{q}_{z} = \int dx d^{2} \mathbf{k}_{\perp} \frac{\mathbf{k}^{2}_{\perp}}{M^{2}} G_{1,1} (x,0,\mathbf{k}^{2}_{\perp},0,0).
\end{equation} 
We consider here the TMD limit, i.e., $\Delta = 0$, which reduces the GTMD to a function of only $x$ and $\mathbf{k}_{\perp}$. The GTMDs are connected to the Wigner correlator as follows \cite{Kaur:2019jow}
\begin{align} \label{wigner}
    \hat{W}^{[\gamma^{+}]} &= F_{1,1},\\
    \hat{W}^{[\gamma^{+} \gamma_{5}]} &=- \frac{i\epsilon^{ij}_{\perp}\mathbf{k}_{\perp}^{i} \Delta_{\perp}^{i}}{M^{2}} G_{1,1},\\
\hat{W}^{[i\sigma^{j+} \gamma_{5}]} &= - \frac{i\epsilon^{ij}_{\perp}\mathbf{k}_{\perp}^{i}}{M^{2}} H_{1,1} - \frac{i\epsilon^{ij}_{\perp} \Delta_{\perp}^{i}}{M^{2}} H_{1,2}.
\end{align}
Here $\epsilon_{\perp}^{ij} = \epsilon^{-+ij}$ is the anti-symmetric tensor, $\epsilon^{0123} = 1$ and $\sigma^{ab} = \frac{i}{2} \left [\gamma^{a}, \gamma^{b}  \right ]$. The Wigner correlator is denoted by the symbol $W^{[\Gamma]}$ and can be expressed as 
\begin{align}
\mathrm{W}^{[\Gamma]} (x,P,  \Delta, \mathbf{k}_{\perp}) &= \frac{1}{2} Tr [ W(x,P, \Delta, \mathbf{k}_{\perp}) \Gamma] \\
&=  \frac{1}{2} \int\frac{dz^{-} d^{2}z_{\perp}}{2 (2\pi)^{3}} e^{ik.z} \langle p' \vert \overline{\psi} \left (-\frac{z^{-}}{2} \right) \Gamma \mathcal{W} \psi \left (\frac{z^{-}}{2} \right) \vert p \rangle \vert_{z^{+} = 0}.
\end{align}
Here, $W^{[\Gamma]}(x,P, \Delta, \mathbf{k}_{\perp})$ is the generalized parton correlation function (GPCF) of the meson. $\mathcal{W}$ refers to the Wilson lines which result from the parallel transit of gauge variables across closed loops. To simplify our present calculations, we consider $\mathcal{W}$ to be equal to 1. $\Gamma$ is the operator sandwiched between the initial and final meson states ($p$ and $p'$) respectively. The GTMDs can be obtained from the GPCFs by integrating over the quark momentum $\mathbf{k}_{\perp}$ \cite{Echevarria:2016mrc}. $F_{1}$ and $H_{1}$ are the GPD limits of the more general GTMDs $F_{1,1}$ and $G_{1,1}$. However, the GTMD $G_{1,1}$ does not have an equivalent GPD due to its $\mathbf{k}_{\perp}$-odd property \cite{Tan:2021osk}. Therefore, the relation stated in Eq. (\ref{eq:4.10}) provides an alternative formulation for the SOC based on a broader parton correlation structure.
In the overlap representation, the leading-twist generalized correlator for GTMDs can be expressed as \cite{Ma:2018ysi,Diehl:2000xz} 

\begin{eqnarray} \label{wignerreal}
    W^{[\gamma^{+}]} &=& \frac{1}{16\pi^{3}} \sum_{\lambda_{\overline{q}}} \left(\phi_{\downarrow\lambda_{\overline{q}}}^\ast (x^{''}, \mathbf{k}_{\perp}^{''}) \phi_{\downarrow\lambda_{\overline{q}}}(x^{'}, \mathbf{k}_{\perp}^{'})  + (\phi_{\uparrow\lambda_{\overline{q}}}^\ast (x^{''}, \mathbf{k}_{\perp}^{''}) \phi_{\uparrow\lambda_{\overline{q}}}(x^{'}, \mathbf{k}_{\perp}^{'})\right), \\
    W^{[\gamma^{+}\gamma_{5}]} &=& \frac{1}{16\pi^{3}} \sum_{\lambda_{\overline{q}}} \left(\phi_{\uparrow\lambda_{\overline{q}}}^\ast (x^{''}, \mathbf{k}_{\perp}^{''}) \phi_{\uparrow\lambda_{\overline{q}}}(x^{'}, \mathbf{k}_{\perp}^{'}) - \phi_{\downarrow\lambda_{\overline{q}}}^\ast(x^{''}, \mathbf{k}_{\perp}^{''}) \phi_{\downarrow\lambda_{\overline{q}}}(x^{'}, \mathbf{k}_{\perp}^{'}) \right), \label{wignerreal1} \\ 
    W^{[i \sigma^{j+}\gamma_{5}]} &=& \frac{1}{16\pi^{3}} \sum_{\lambda_{\overline{q}}} \left(\phi_{\uparrow\lambda_{\overline{q}}}^\ast (x^{''}, \mathbf{k}_{\perp}^{''}) \phi_{\uparrow\lambda_{\overline{q}}}(x^{'}, \mathbf{k}_{\perp}^{'}) - \phi_{\downarrow\lambda_{\overline{q}}}^\ast(x^{''}, \mathbf{k}_{\perp}^{''}) \phi_{\downarrow\lambda_{\overline{q}}}(x^{'}, \mathbf{k}_{\perp}^{'}) \right). \label{wignerreal2}
\end{eqnarray}
The arguments of the initial-state wave functions in Eqs. (\ref{wignerreal}), (\ref{wignerreal1}) and (\ref{wignerreal2}) are given as
\begin{align*}
x_{1} &= \frac{x - \zeta/2}{1 - \zeta/2}, \\
k_{\perp 1} &= k_{\perp} - \frac{1- x}{1 - \zeta/2} \frac{\Delta_{\perp}}{2},
\end{align*} 
and for the final-state wave functions, they are given as 
\begin{align*}
    x_{2} &= \frac{x + \zeta/2}{1 + \zeta/2}, \\
    k_{\perp 2} &= k_{\perp} + \frac{1- x}{1 + \zeta/2} \frac{\Delta_{\perp}}{2}.
\end{align*}

\subsection{Spin-0 mesons}

The leading-twist GTMDs that we mention in this work are $F_{1,1}$,  $G_{1,1}$, $H_{1,1}$ and $H_{1,2}$. Using the LFWFs of the form of Eq. (\ref{3.13}), along with the overlap representation for $W^{[\Gamma]}$, we obtain the explicit expressions for the GTMDs of mesons having disparate quark and anti-quark masses \cite{Kaur:2019jow} 
\begin{align}
    F_{1,1} &= \frac{1}{16\pi^{3}} \left [ \mathbf{k}_{\perp}^{2}  + \mathcal{M}_{2}  \mathcal{M}_{1}\right ] \times \frac{\phi (x_{1}, \mathbf{k}_{\perp 1})\phi^{\dagger} ( x_{2}, \mathbf{k_{\perp2}})}{\sqrt{ j^{2}_{2} + \mathbf{k}_{\perp 2}^{2}} \sqrt{j^{2}_{1} + \mathbf{k}_{\perp 1}^{2}}},\\
    G_{1,1} &= - \frac{M^{2}}{16\pi^{3}} \frac{(2 -  x_{2} - x_{1})}{2}  \frac{\phi (x_{1}, \mathbf{k}_{\perp 1})\phi^{\dagger} ( x_{2}, \mathbf{k}_{\perp 2})}{\sqrt{ j^{2}_{2} + \mathbf{k}_{\perp 2}^{2}} \sqrt{j^{2}_{1} + \mathbf{k}_{\perp 1}^{2}}}, \\
     H_{1,1} &= - \frac{M}{16\pi^{3}} \left [   \mathcal{M}_{1} - \mathcal{M}_{2}\right ]  \frac{\phi (x_{1}, \mathbf{k}_{\perp 1})\phi^{\dagger} ( x_{2}, \mathbf{k}_{\perp 2})}{\sqrt{ j^{2}_{2} + \mathbf{k}_{\perp 2}^{2}} \sqrt{j^{2}_{1} + \mathbf{k}_{\perp 1}^{2}}},\\
     H_{1,2} &= \frac{M}{16\pi^{3}} \left [  \mathcal{M}_{1} \frac{(1 -  x_{2})}{2} - \mathcal{M}_{2} \frac{(1 -  x_{1})}{2}\right ]  \frac{\phi (x_{1}, \mathbf{k}_{\perp 1})\phi^{\dagger} ( x_{2}, \mathbf{k}_{\perp 2})}{\sqrt{ j^{2}_{2} + \mathbf{k}_{\perp 2}^{2}} \sqrt{j^{2}_{1} + \mathbf{k}_{\perp 1}^{2}}},
\end{align}
where
\begin{eqnarray}    
    \mathcal{M}_{1} &=& \frac{1 - x}{1 + \zeta} m_{q} + \frac{x + \zeta}{1 + \zeta} m_{\overline{q}}, \\
    \mathcal{M}_{2} &=& \frac{1 - x}{1 - \zeta} m_{q} + \frac{x - \zeta}{1 - \zeta}m_{\overline{q}}, \\
    j^{2}_{1} &=& \frac{1 - x}{1 + \zeta} m_{q}^{2} + \frac{x + \zeta}{1 + \zeta} m_{\overline{q}}^{2} - \frac{(1-x)(x + \zeta)}{(1 + \zeta^2)^{2}} (m_{q} - m_{\overline{q}})^{2}, \\
        j^{2}_{2} &=& \frac{1 - x}{1 - \zeta} m_{q}^{2} + \frac{x - \zeta}{1 - \zeta} m_{\overline{q}}^{2} - \frac{(1-x)(x - \zeta)}{(1 - \zeta^2)^{2}} (m_{q} - m_{\overline{q}})^{2}.
\end{eqnarray}

In this section, we have presented the quark GTMDs of mesons with respect to the longitudinal momentum fraction carried by quark $x$. Being the mother distributions, GTMDs have the versatility to be reduced to the corresponding GPDs and TMDs. The $\mathbf{k}_{\perp}$-even GTMDs are reduced to the respective GPDs after integrating over $\mathbf{k}_{\perp}$ \cite{Tan:2021osk}. We have

\begin{eqnarray}
    F_{1}^{q} (x, \zeta, t) &=& \int d^{2}\mathbf{k}_{\perp} F_{1,1}, \\
    H_{1}^{q} (x, \zeta, t) &=& \int d^{2}\mathbf{k}_{\perp} \left( \frac{\mathbf{k}_{\perp}.\Delta_{ \perp}}{\Delta^{2}_{\perp}} H_{1,1} + H_{1,2} \right ).
\end{eqnarray}
Further, the anti-quark GTMDs are related to the quark GTMDs by the relation
\begin{equation}
    F^{u} (x, \mathbf{k}_{\perp}^{2}, \zeta, \Delta^{2}_{\perp}, \mathbf{k}_{\perp}.\Delta_{\perp}, m_{q}, m_{\overline{q}}) =  F^{\overline{s}} (1-x, \mathbf{k}_{\perp}^{2}, \zeta, \Delta^{2}_{\perp}, -\mathbf{k}_{\perp}.\Delta_{\perp}, m_{\overline{q}},m_{q}).
\end{equation}

\subsection{Spin-1 mesons}

For spin-1 mesons, our spatial wave function will remain the same but there will be an addition of a spin wave part. Since we are essentially dealing with TMDs, in this section we define the explicit expression of $g_{1} (x, \mathbf{k}_{\perp}^{2})$   T-even TMDs \cite{Puhan:2023hio,Kaur:2020emh} in the LFQM using the wave functions in Eq. (\ref{3.13}). We have
\begin{align} \label{eq4.30}
	g_{1}\left(x, \mathbf{k}_{\perp}^{2}\right) &=\frac{M}{2(2 \pi)^{3}}\left( 2 m_{q} + M\right) \left[(2 \mathbf{k}_{\perp}^{2}+m_{q}\left(M + 2 m_{q} \right)) +  m_{q} M(1-2 x)\right] \\
 &\times\frac{\left|\psi\left(x,\mathbf{k}_{\perp}^{2}\right)\right|^{2}}{\omega^{2}},
\end{align}
where
\begin{equation}
    \omega = (M + 2 m_{q})\sqrt{\mathbf{k}_{\perp}^{2} + m_{q}^{2}}.
\end{equation}

\section{NUMERICAL RESULTS}{\label{sec5}}

For the numerical calculations, we have taken the input parameters of the LFQM for different quark masses $(m_{q}, m_{b}, m_{c}, m_{s})$ with $q = (u,d)$ and different variational HO parameters ($\beta_{q\overline{q}},  
  \beta_{q\overline{c}},  \beta_{q\overline{b}},   \beta_{q\overline{s}}, 
   \beta_{s\overline{c}},   \beta_{b\overline{b}}, \beta_{c\overline{c}}, 
     \beta_{s\overline{b}}, \beta_{c\overline{b}}$) from Ref. \cite{Arifi:2022pal, Tan:2021osk}. These parameters have been presented in Table \ref{tab:i} and obtained by reproducing the mass spectra using the variational principle which has been successful in computing various physical properties such as decay constants, electromagnetic form factors and distribution amplitudes \cite{Arifi:2022pal}. 

\begin{table}[htbp] {\label{T1}}
\centering
\footnotesize
\begin{tabular}{|c|c|c|c|c|c|c|c|c|c|c|c|c|}
\hline
$m_{q}$ & $m_{c}$ & $m_{s}$ & $m_{b}$ & $\beta_{q\overline{q}}$ & $\beta_{q\overline{c}}$ &  $\beta_{q\overline{b}}$ &   $\beta_{q\overline{s}}$ & $\beta_{s\overline{c}}$ &   $\beta_{b\overline{b}}$ & $\beta_{c\overline{c}}$ & $\beta_{s\overline{b}}$ & $\beta_{c\overline{b}}$\\
\hline
0.22&1.68&0.45&5.10&0.523&0.500&0.585&0.524&0.537&1.376&0.699&0.636&0.906\\
\hline
\end{tabular}
\caption{Model Parameters for LFQM\label{tab:i}.}
\end{table}
In order to compute the spin-orbit correlators using the $G_{1,1}$ GTMD for the respective meson, we have used Eq. (\ref{eq:4.10}) along with the quark masses and HO parameters from Table \ref{tab:i}. We have summarized the calculated results of the SOC inside various spin-0 mesons in Table \ref{table2}.
From the table we observe that the sign of the correlation is negative for all the spin-0 mesons which clearly implies that the quark longitudinal spin and quark OAM tend to be anti-aligned inside the respective spin-0 meson. This correlation between the quark spin and OAM takes into account the effective number of quarks inside a parent hadron \cite{Lorce:2011kd}. Therefore, it would be interesting to compare the absolute value of the spin-0 mesons to that of the nucleons. It is clear from the results that the magnitude of SOC for the mesons is less than that of the nucleon ($C_{z}^{u/n} = -0.9$ and $C_{z}^{d/n} = -0.53$ \cite{Lorce:2014mxa}) pointing towards a weaker correlation inside the mesons which seems to be due to more number of effective quarks inside the nucleons in comparison to those in the mesons. 
\begin{table}[htbp]
	\centering
	\begin{tabular}{|c|c|}
		\hline
		spin-0 mesons  & $C_{z}^{q/M}$ \\
		\hline
		$\pi^{+}$ & $-$0.272 \\
		$K^{+}$ & $-$0.251 \\
		$K^{0}$ & $-$0.251 \\
		$B^{+}$ & $-$0.227 \\
		$B^{0}$ &  $-$0.227 \\
		$B^{0}_{s}$ & $-$0.161 \\
		$B^{+}_{c}$ & $-$0.035 \\
		$D^{+}$ & $-$0.082 \\
		$D^{0}$ & $-$0.082 \\
		$D^{+}_{s}$ & $-$0.072\\
		$\eta_{b}$ & $-$0.031\\
		$\eta_{c}$ & $-$0.063 \\
		\hline
	\end{tabular}
	\caption{Spin-orbit correlation $C_{z}^{q}$ for spin-0 mesons. }
	\label{table2}
	\hfill
\end{table}

\begin{figure}[htbp]
\centering
\begin{subfigure}[b]{0.45\textwidth}
    \includegraphics[width=\textwidth]{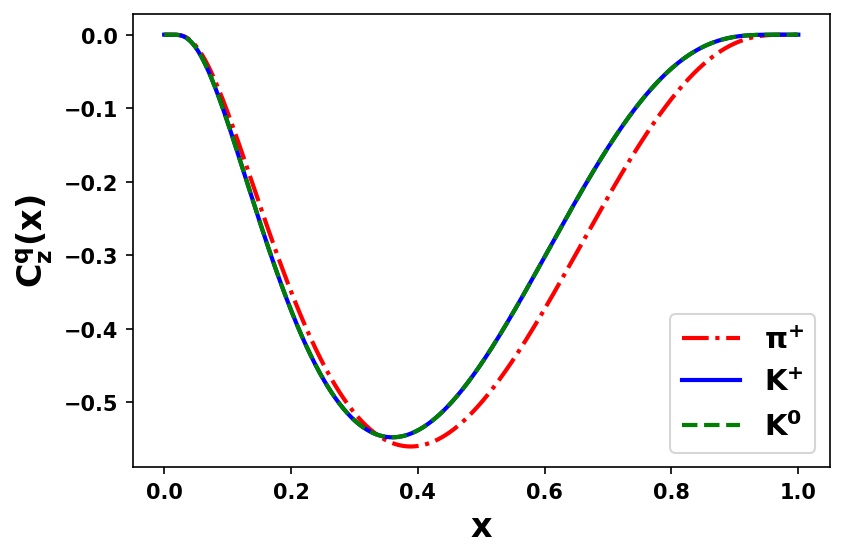}
     \caption{} \label{(a1)}
\end{subfigure}
\begin{subfigure}[b]{0.45\textwidth}
    \includegraphics[width=\textwidth]{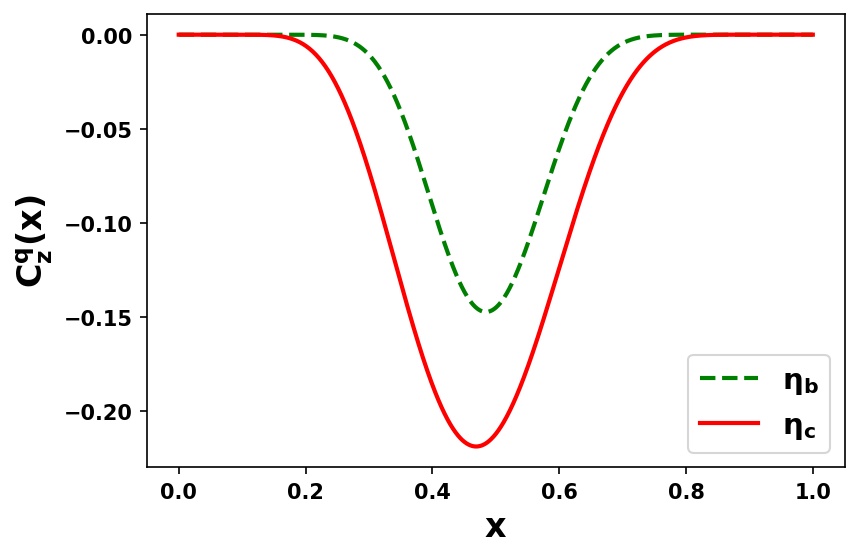}
    \caption{} \label{(b1)}
\end{subfigure}
\qquad
\begin{subfigure}[b]{0.45\textwidth}
    \includegraphics[width=\textwidth]{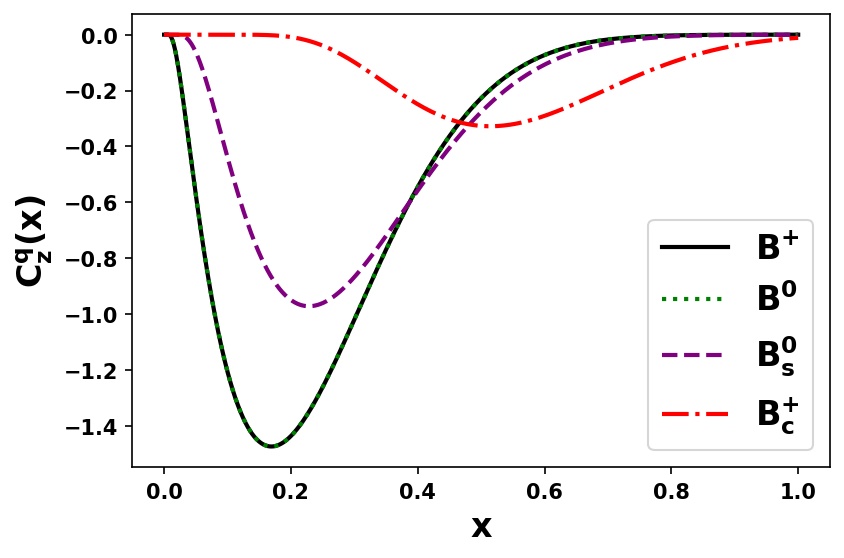}
    \caption{} \label{(c1)}
\end{subfigure}
\begin{subfigure}[b]{0.45\textwidth}
    \includegraphics[width=\textwidth]{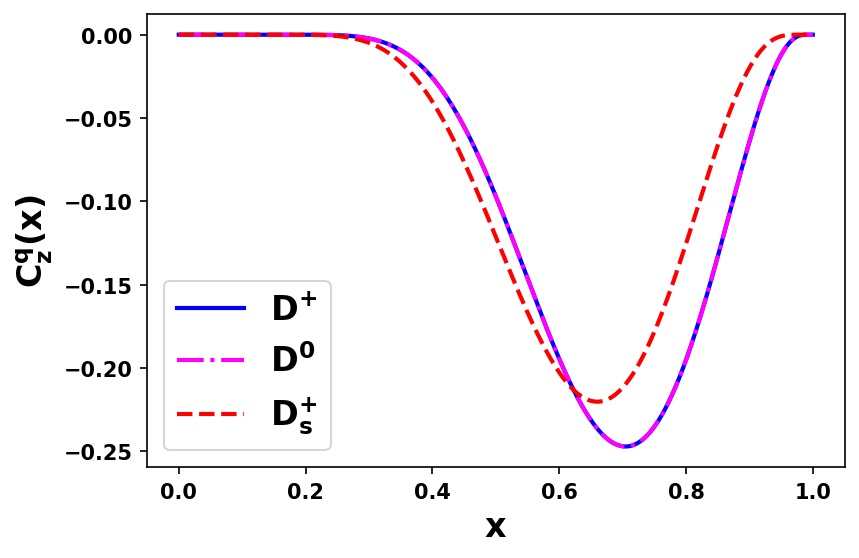}
    \caption{} \label{(d1)}
\end{subfigure}
\caption{$x$-dependence of  $C_{z}^{q} (x)$ for all spin-0 mesons\label{fig:1}.}
\end{figure}

In order to show the dependence of quark SOC on the range of longitudinal momentum fraction $x$, we integrate $C_{z}^{q}$ over $\mathbf{k}_{\perp}$ and show the variation of $C_{z}^{q} (x)$ with respect to $x$ for various spin-0 mesons in Fig. \ref{fig:1}.  In Fig. \ref{fig:1}(a), we present the most commonly studied mesons: the pion and the kaon. Here, it is observed that the largest contribution for the pion and kaon comes from the region where the longitudinal momentum fraction $x$ is around $0.4$ and $0.38$ respectively. In Fig. \ref{fig:1}(b), we present the $\eta_{b}$ and $\eta_{c}$ mesons and their highest $x$ contribution comes at $0.5$ and $0.5$ respectively. 
The plots in this case are symmetric which is due to negligible difference in quark and anti-quark masses. However, the $C_{z}^{q}$ values for $\eta_{b}$ and $\eta_{c}$ are extremely low when compared to those of pion and kaon. 
Similarly, in Fig. \ref{fig:1}(c), we have considered the $B$-mesons. The largest contribution for the $B$-mesons $B^{+}$, $B^0$, $B^{0}_{s}$, and $B^{+}_{c}$ comes from regions where $x$ is around $0.18$, $0.18$, $0.21$, and $0.5$, respectively. The peaks shift towards higher $x$ values which is due to the increasing inequality in the quark and anti-quark masses inside the $B$-mesons having the quark contents as $B^{+} (u \bar{b})$, $B^0 (d \bar{b})$, $B^{0}_{s} (s \bar{b})$, and $B^{+}_{c} (c \bar{b})$. For the case where the quark is lighter than the other anti-quark in the meson, a smaller longitudinal momentum fraction $x$ is carried by the quark hence leading to the distribution peak at lower values of $x$. Further, in Fig. \ref{fig:1}(d) all of the $D$-mesons have been presented. It has been found that the distribution peak for the $D$-mesons $ D^{+}$, $ D^{0}$ and $D^{+}_{s}$ for the longitudinal momentum fraction $x$ at $0.7$, $0.7$ and $0.62$ respectively. For $ D^{+}$ and $ D^{0}$, the quark ($c$) being heavier than the anti-quark ($u$ or $d$) carries a larger longitudinal momentum fraction $x$. This shifts the peak of the distribution to higher values of $x$ and the curve is shifted to the right.
For the case of $D^{+}_{s}$, the difference between the quark ($c$) and the anti-quark ($s$) is less as compared to that of $ D^{+}$ or $ D^{0}$, the distribution peaks at a comparatively lower $x$ value.  

\begin{figure}[htbp]
	\centering
	\begin{subfigure}[b]{0.45\textwidth}
		\includegraphics[width=\textwidth]{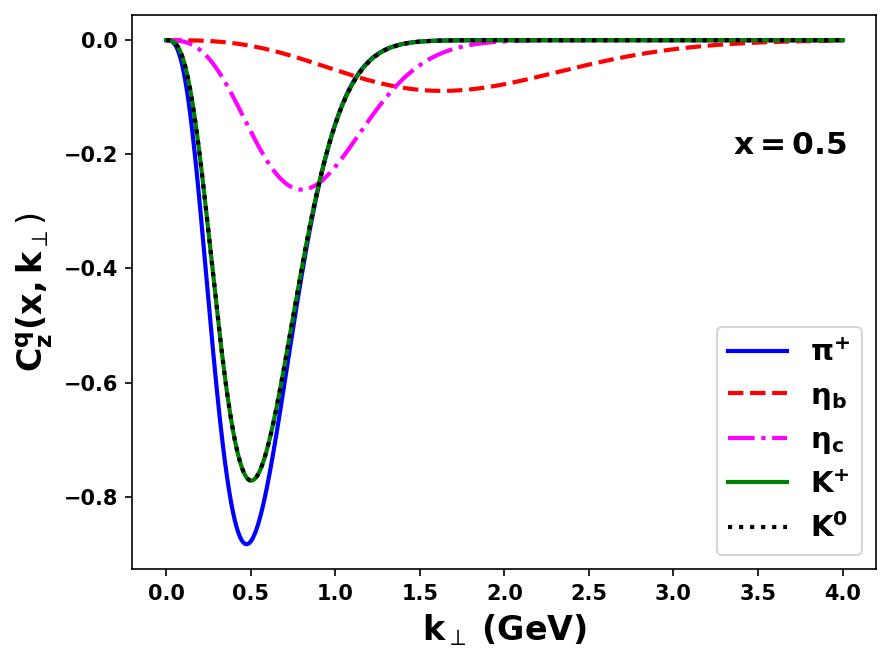}
		\caption{} \label{(a2)}
	\end{subfigure}
	\begin{subfigure}[b]{0.45\textwidth}
		\includegraphics[width=\textwidth]{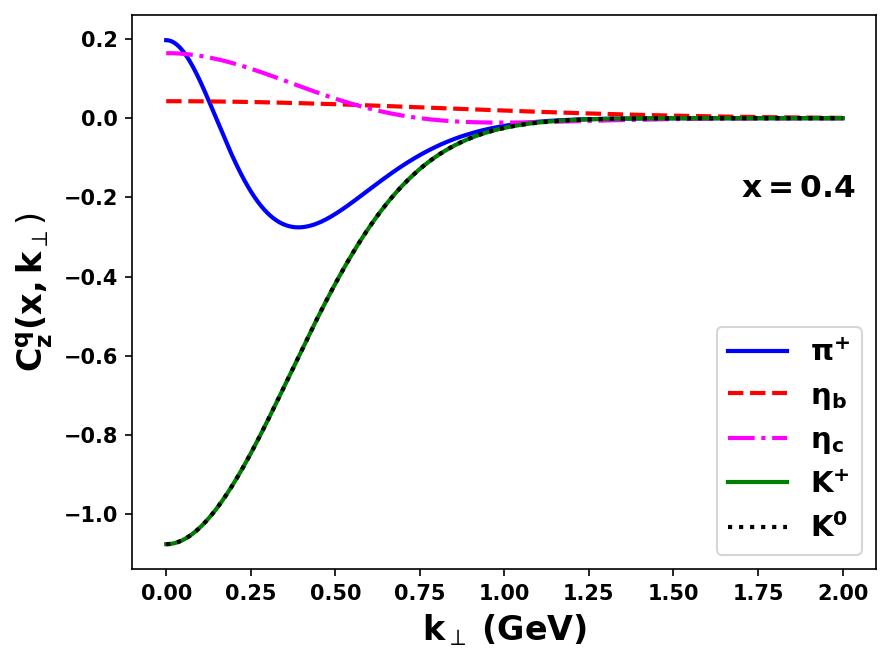}
		\caption{} \label{(b2)}
	\end{subfigure}
	\begin{subfigure}[b]{0.45\textwidth}
		\includegraphics[width=\textwidth]{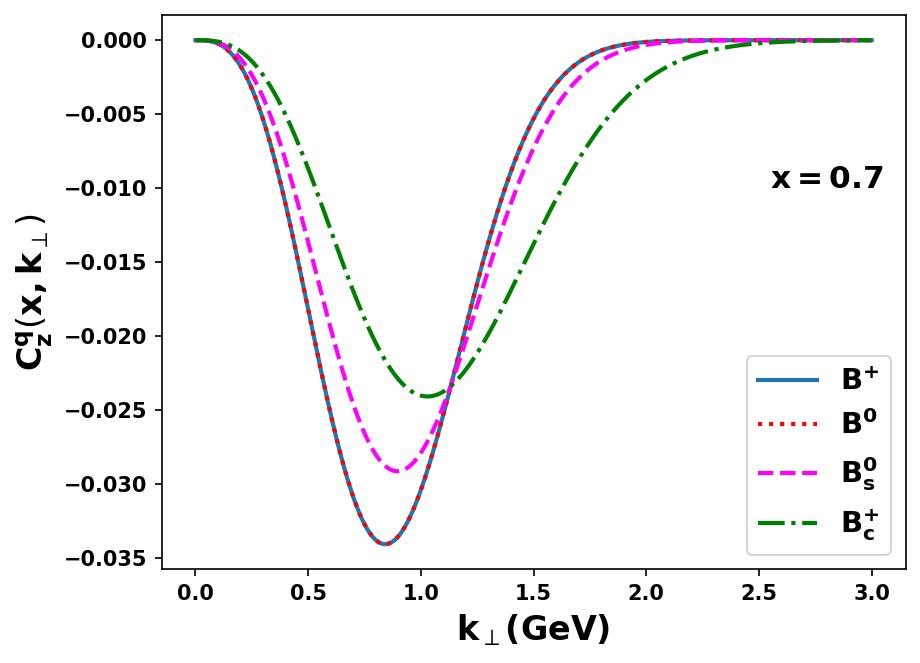}
		\caption{} \label{(c2)}
	\end{subfigure}
	\begin{subfigure}[b]{0.45\textwidth}
		\includegraphics[width=\textwidth]{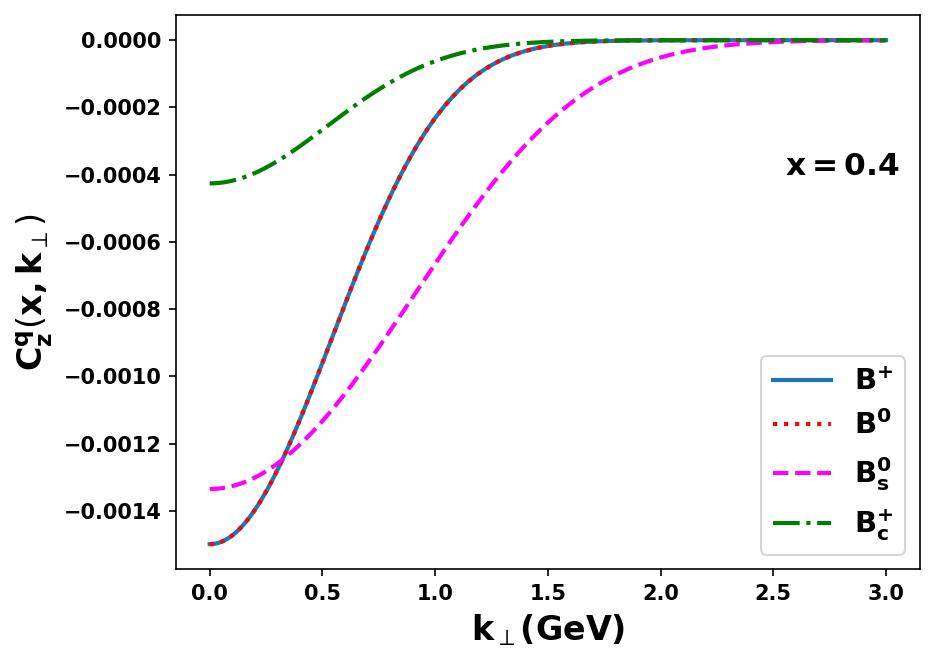}
		\caption{} \label{(d2)}
	\end{subfigure}
	\begin{subfigure}[b]{0.45\textwidth}
		\includegraphics[width=\textwidth]{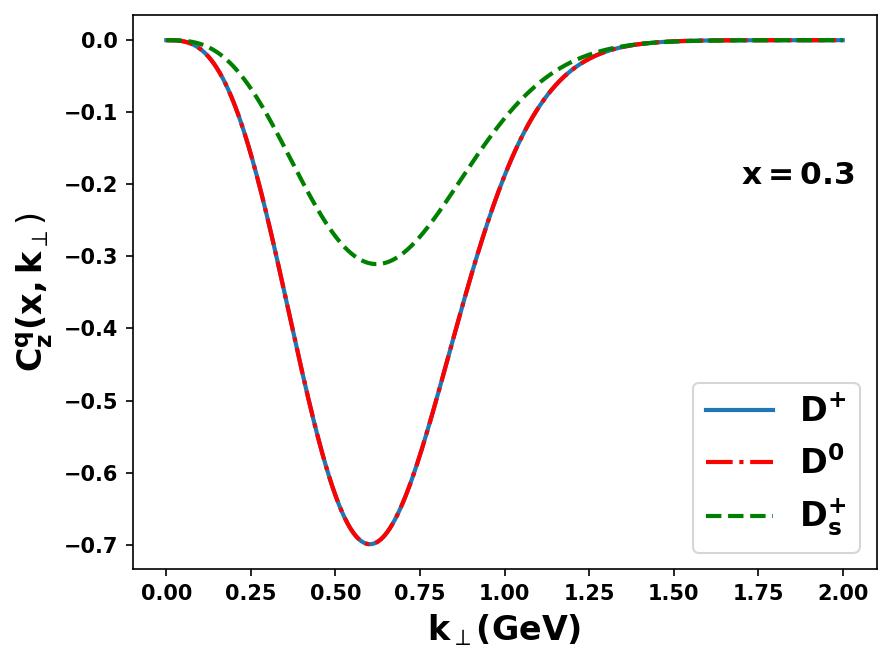}
		\caption{} \label{(e2)}
	\end{subfigure}
	\begin{subfigure}[b]{0.45\textwidth}
		\includegraphics[width=\textwidth]{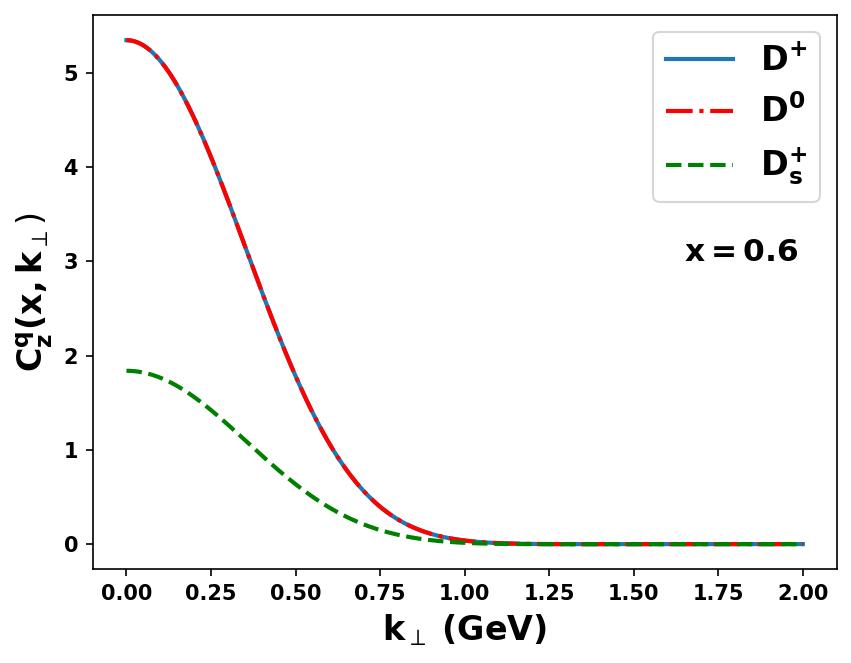}
		\caption{} \label{(f2)}
	\end{subfigure}
	\caption{$\mathbf{k}_{\perp}$ dependence of SOC $C_{z}^{q} (x, \mathbf{k}_{\perp})$ at a fixed longitudinal momentum fraction $x$ for spin-0 mesons. The left panel shows the dependence of SOC on the $\mathbf{k}_{\perp}$ using the GTMD approach whereas the right panel  shows the dependence using the GPD approach.\label{fig:2}}
\end{figure}

Further, in Fig. \ref{fig:2} we have presented the dependence of spin-orbit contribution with transverse momentum $\mathbf{k}_{\perp}$  at fixed values of longitudinal momentum fraction $x$. 
The left panel in Fig. \ref{fig:2} (\ref{fig:2}(a), \ref{fig:2}(c) and \ref{fig:2}(e)) shows the dependence of SOC on the $\mathbf{k}_{\perp}$ using the GTMD approach whereas the right panel of Fig. \ref{fig:2} (\ref{fig:2}(b), \ref{fig:2}(d) and \ref{fig:2}(f)) shows the dependence using the GPD approach. Fig. \ref{fig:2}(a) presents the variation of $\pi^{+}$, $K^{+}$, $K^{0}$, $\eta_b$, $\eta_c$ at $x=0.5$, Fig. \ref{fig:2}(c) presents $B^{+}$, $B^{0}$, $B^{0}_{s}$, $B^{+}_{c}$ at $x=0.7$ and Fig. \ref{fig:2}(e) presents $D^{+}$, $D^{0}$, $D^{+}_{s}$ at $x=0.3$. On the other hand, Fig. \ref{fig:2}(b) presents the variation of $\pi^{+}$, $K^{+}$, $K^{0}$, $\eta_b$, $\eta_c$ at $x=0.4$, Fig. \ref{fig:2}(d) presents $B^{+}$, $B^{0}$, $B^{0}_{s}$, $B^{+}_{c}$ at $x=0.4$ and Fig. \ref{fig:2}(f) presents $D^{+}$, $D^{0}$, $D^{+}_{s}$ at $x=0.6$. It would be important to mention here that we have taken different values of $x$ for each set to project the difference between different mesons. Keeping the same $x$ values will not affect the dependence but will only affect the amplitude of the SOC. It is observed that even though we have the same $C_{z}^{q}$ value for both the GTMD and GPD approaches, the transverse momentum dependence is different for them \cite{Tan:2021osk}. For the GTMD case, the $C_{z}^{q} (x, \mathbf{k}_{\perp})$ is negative over the whole region of  $\mathbf{k}_{\perp}$ for all the mesons which is in agreement with pion case detailed in Ref. \cite{Tan:2021osk}. We notice that for the mesons with light quarks, the peaks occur at lower values of $\mathbf{k}_{\perp}$ but with a comparatively large  amplitude. The peaks are narrow and sharp for the light quark mesons but as we increase the value of $\mathbf{k}_{\perp}$, they diminish and tend to zero. For the case of heavy quark mesons, the peaks appear at higher values of transverse momentum and are broader. The  amplitudes also become quite small. 
Further, using the GPD approach, $C_{z}^{q} (x, \mathbf{k}_{\perp})$ comes out to be positive for higher values of $\mathbf{k}_{\perp}$  in the case of  $\pi^{+}$, $K^{+}$, $K^{0}$, $\eta_b$, $\eta_c$ presented in Fig. \ref{fig:2}(b) for $x=0.4$. As the $\mathbf{k}_{\perp}$ values decrease, the $C_{z}^{q} (x, \mathbf{k}_{\perp})$ value first decreases and then increases for the case of $\pi^{+}$. For $\eta_c$, it increases with decreasing $\mathbf{k}_{\perp}$, for  $K^{+}$ and $K^{0}$, it decreases with decreasing $\mathbf{k}_{\perp}$ and for $\eta_b$ there is a negligible increase in value   with decreasing $\mathbf{k}_{\perp}$. 
These results are in agreement with the results in Ref. \cite{Tan:2021osk}. 
For the case of $B$-mesons $B^{+}$, $B^{0}$, $B^{0}_{s}$, $B^{+}_{c}$ in Fig. \ref{fig:2}(d), the variation of $C_{z}^{q} (x, \mathbf{k}_{\perp})$ has been presented for $x=0.4$. In these cases, the results are negative throughout the $\mathbf{k}_{\perp}$ region but tend to zero for higher values of $\mathbf{k}_{\perp}$. The SOCs for the $D$-mesons presented in Fig. \ref{fig:2}(f) for $x=0.6$ remain positive but tend to zero beyond $\mathbf{k}_{\perp}$=1.00 GeV. This opposite behavior of the $B$-mesons and $D$-mesons is due to the difference in the quark distributions having light and heavy masses respectively.

We now compute the SOC for the case of spin-1 mesons. We consider  Eq. (\ref{eq:4.10}) and replace the $G_{1,1}$ relation with the expression $g_{1} (x,\mathbf{k}_{\perp})$ TMD from Eq. (\ref{eq4.30}). The numerical results of $C_{z}^{q}$ for the spectrum of spin-1 mesons having definite quark contents have been presented in Table \ref{tab:my_label}. The sign of the correlation comes out to be  positive for spin-1 mesons implying that the quark OAM and the quark longitudinal spin tend to be directly aligned inside the respective spin-1 mesons. The difference between the alignment of SOC for the spin-1 meson and the spin-0 meson is because of the spin density term in the energy-momentum tensor defined in Eq. (\ref{momtensor}). There are only two possibilities of the  value of $C_{z}^{q}$ which can be either  positive or negative depending on the alignment of the quark longitudinal spin and quark OQM. For the case of spin-1 mesons,  positive values are obtained.

\begin{table}[h]
    \centering
    \begin{tabular}{|c|c|}
    \hline
       spin-1 mesons  & $C_{z}^{q/M}$ \\
       \hline
        $\rho^{+}$ & 0.332 \\
         $J/\psi$ & 0.241 \\
         $\Upsilon$ & 0.338\\
         $\phi$ & 0.221 \\
         $K^{*+}$ & 0.291 \\
         $K^{*0}$ & 0.291 \\
         $B^{*0}$ & 0.179 \\
         $B^{*+}$ & 0.179 \\
         $B^{*0}_{s}$ & 0.191 \\
         $B^{*+}_{c}$ & 0.224 \\
         $D^{*0}$ & 0.411 \\
         $D^{*+}$ & 0.411 \\
         $D^{*+}_{s}$ & 0.336\\
         \hline
    \end{tabular}
    \caption{Spin-orbit correlation $C_{z}^{q}$ for spin-1 mesons. }
    \label{tab:my_label}
\end{table}

\begin{figure}[!]
\centering
\begin{subfigure}[b]{0.49\linewidth}
    \includegraphics[width=\linewidth]{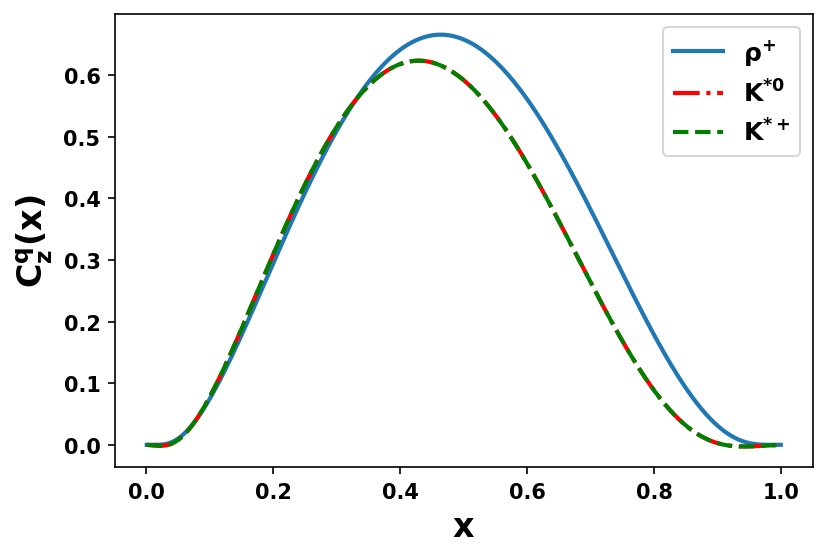}
     \caption{} \label{(a3)}
\end{subfigure}
\begin{subfigure}[b]{0.49\linewidth}
   \includegraphics[width=\linewidth]{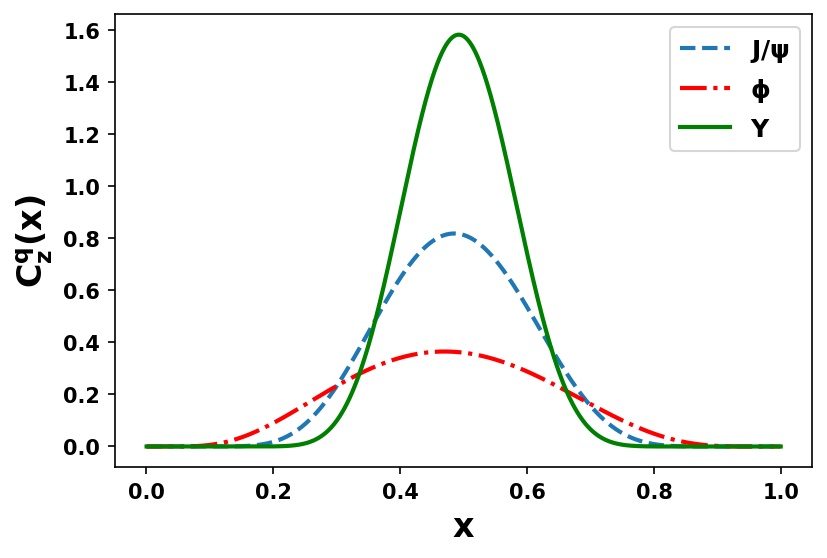}
     \caption{} \label{(b3)}
\end{subfigure}
\begin{subfigure}[b]{0.49\linewidth}
    \includegraphics[width=\linewidth]{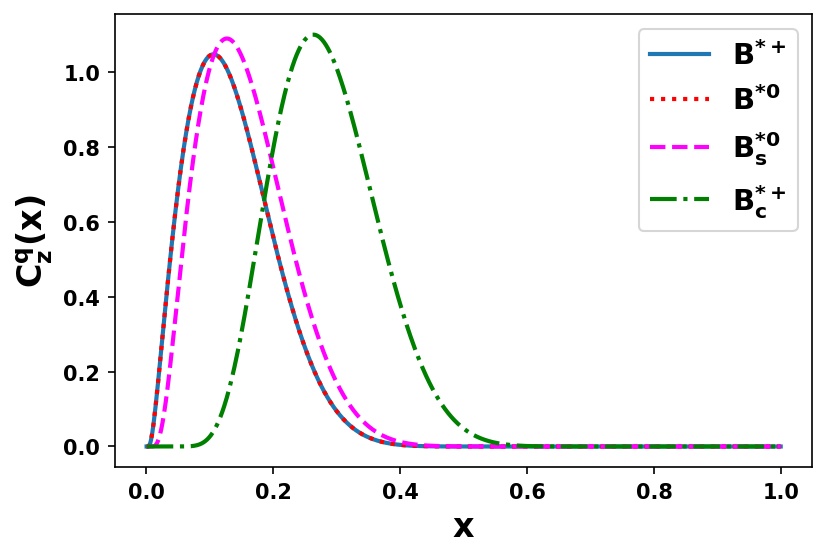}
     \caption{} \label{(c3)}
\end{subfigure}
\begin{subfigure}[b]{0.49\linewidth}
    \includegraphics[width=\linewidth]{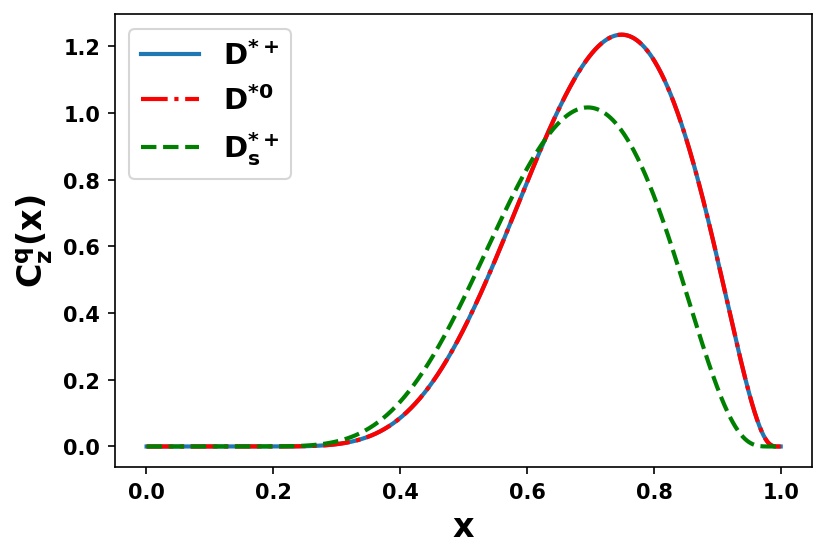}
     \caption{} \label{(d3)}
\end{subfigure}
\caption{$x$-dependence of $C_{z}^{q}(x)$ for all spin-1 mesons\label{fig:3}.}
\end{figure}

In Fig. \ref{fig:3} we have presented the dependence of $C_{z}^{q}$ on longitudinal momentum fraction $x$ for different spin-1 mesons. We discuss the cases of $\rho^{+}$, $K^{*0}$ and $K^{*+}$ in Fig. \ref{fig:3}(a) where the maximum contribution of $C_{z}^{q}(x)$ for $\rho^{+}$, $K^{*0}$ and $K^{*+}$ is at $0.50$, $0.48$ and $0.48$  respectively.  In Fig. \ref{fig:3}(b), we present the SOCs for  $J/\psi$, $\phi$ and $\Upsilon$ mesons. In this case a symmetry is observed because of a similar quark content in the mesons. The largest contributions come approximately around $0.50$ for all the mesons in this plot. In Fig. \ref{fig:3}(c), we consider the $B$-mesons and the largest contribution for $ B^{*+}$, $ B^{*0}$, $B^{*0}_{s}$, and $B^{*+}_{c}$ mesons is for the values of $x$ at $0.10$, $0.10$, $0.12$, and $0.30$ for respective mesons. Similar to the case of the spin-0 mesons, in the case on spin-1 mesons the quarks carry a smaller longitudinal momentum fraction $x$ when the quark mass is lighter than its corresponding anti-quark. This results in a peak at lower values of $x$ and the curve shifts towards the left. Finally, in Fig. \ref{fig:3}(d), we present the $D$-mesons. The highest $x$ contribution for the $D$-mesons $ D^{+}$, $ D^{0}$ and $D^{+}_{s}$ is found to be at $x=$ $0.22$, $0.22$ and $0.28$ respectively. The shifting of peak is again due to the difference in the quark and anti-quark masses giving a peak at higher values of longitudinal momentum fraction $x$ when this difference is small.

\begin{figure}[h]
		\centering
		\begin{subfigure}[b]{0.49\linewidth}
			\includegraphics[width=\linewidth]{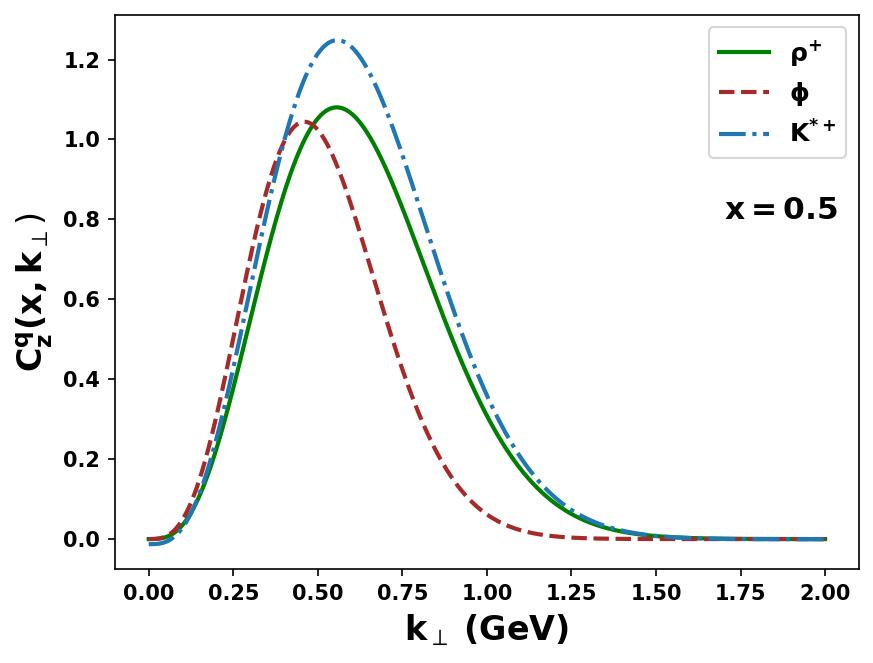}
			\caption{} \label{(a4)}
		\end{subfigure}
		\begin{subfigure}[b]{0.49\linewidth}
			\includegraphics[width=\linewidth]{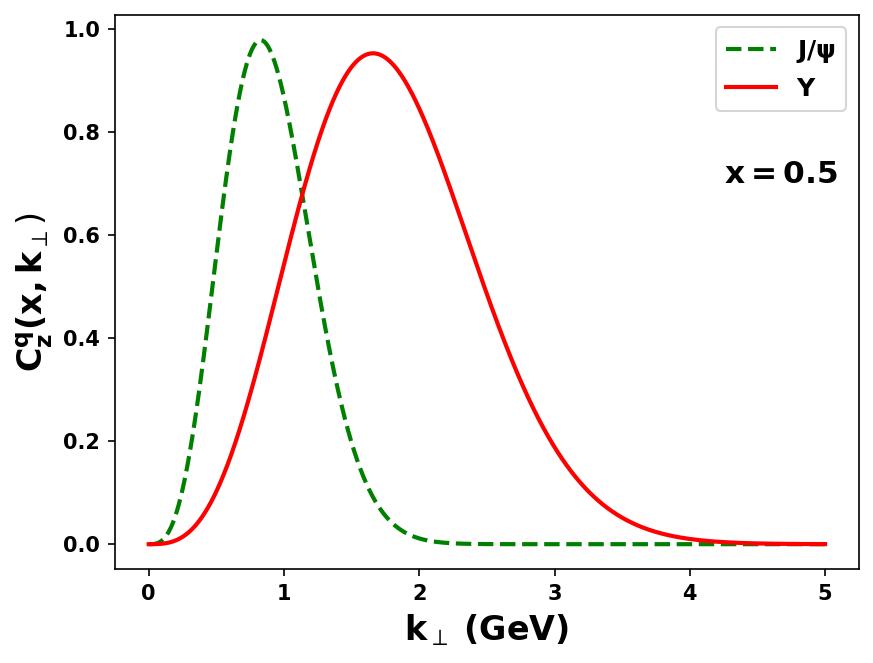}
			\caption{} \label{(b4)}
		\end{subfigure}
		\begin{subfigure}[b]{0.49\linewidth}
			\includegraphics[width=\linewidth]{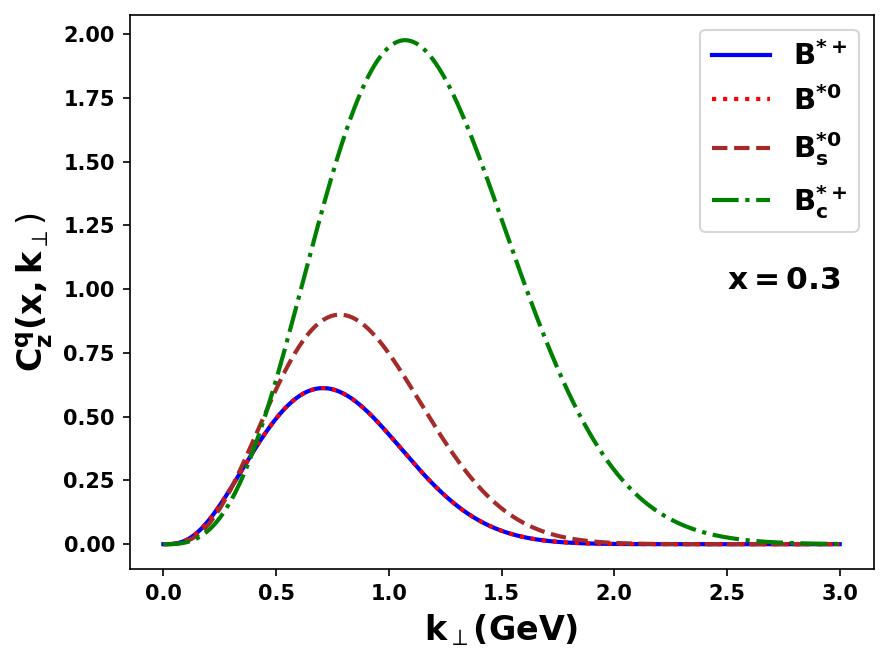}
			\caption{} \label{(c4)}
		\end{subfigure}
		\begin{subfigure}[b]{0.49\linewidth}
			\includegraphics[width=\linewidth]{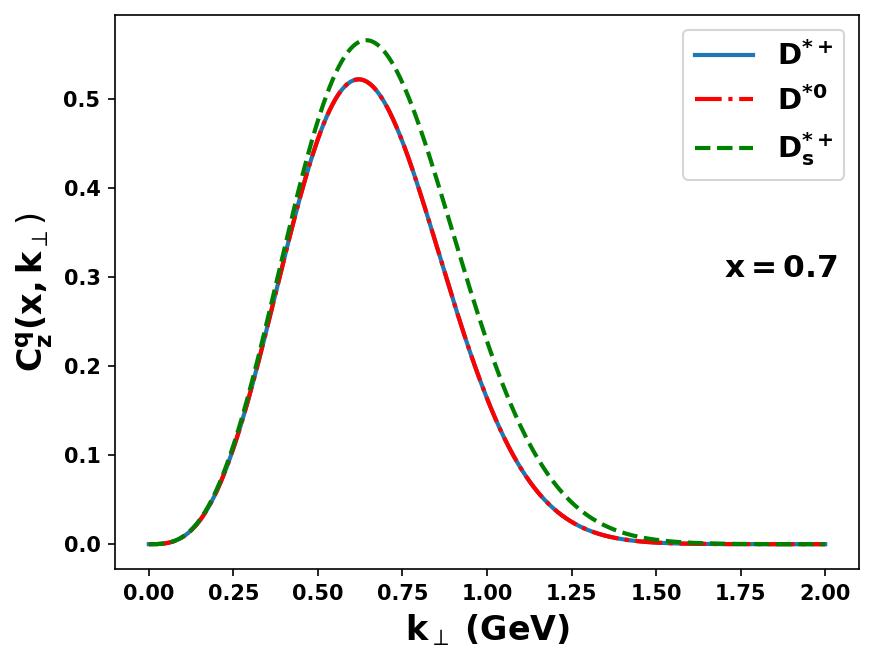}
			\caption{} \label{(d4)}
		\end{subfigure}
		\caption{$\mathbf{k}_{\perp}$-dependence of SOC $C_{z}^{q} (x, \mathbf{k}_{\perp})$ at a fixed longitudinal momentum fraction $x$ for spin-1 mesons using the GTMD approach. \label{fig:4}}
	\end{figure}

In Fig. \ref{fig:4}, we have shown the dependence of spin-orbit contribution with transverse momentum $\mathbf{k}_{\perp}$ at fixed values of longitudinal momentum fraction $x$  for the spin-1 mesons using the GTMD approach. Fig. \ref{fig:4}(a) presents the variation of $\rho^{+}$, $K^{*0}$ and $K^{*+}$ at $x=0.5$, Fig. \ref{fig:4}(b) presents $J/\psi$ and $\Upsilon$  at $x=0.5$, Fig. \ref{fig:4}(c) presents $ B^{*+}$, $ B^{*0}$, $B^{*0}_{s}$, and $B^{*+}_{c}$ at $x=0.3$ and Fig. \ref{fig:4}(d) presents $ D^{+}$, $ D^{0}$ and $D^{+}_{s}$ at $x=0.7$. As discussed  earlier, the peak of the quark distribution depends on the quark content of the meson. Mesons having similar quark and anti-quark masses tend to have a higher amplitude for its distribution. For the case of mesons having a difference in the quark and anti-quark masses, the amplitude varies in proportion to the  mass difference. It is also observed that $C_{z}^{q} (x, \mathbf{k}_{\perp})$ remains positive across the entire $\mathbf{k}_{\perp}$ region.

\section{SUMMARY AND CONCLUSION} {\label{sec6}}
In this work we have studied the correlation between the quark's orbital angular motion and the quark's longitudinal spin inside the light and heavy mesons with spin-0 and spin-1. We started by defining the gauge-invariant LF quark longitudinal OAM and decomposing it into its constituent left-handed and right-handed quark contributions. The quark SOC is described by the difference between the right and left-handed quark contributions of this longitudinal OAM. We defined $\hat{T}^{\mu\nu}_{q5}$ and further decomposed it into two FFs out of which one form factor is the SOC ascertained by the form factor $C_{z}^{q}$. We considered two approaches in order to calculate $C_{z}^{q}$. One is the GTMD approach, where $C_{z}^{q}$ is defined by the leading-twist correlator $G_{1,1}$ for the spin-0 mesons and $g_{1} (x,\mathbf{k}_{\perp})$ for the case of spin-1 mesons. The alternative way is the GPD technique, in which the first $x$ moments of $F_{1}^{q} (x, \zeta, t)$ and $H_{1}^{q} (x, \zeta, t)$ at $\zeta=0$,  $t=0$ $Ge V^{2}$ combine to yield the correlation's expectation value. We calculated the analytical results for $C_{z}^{q}$ by considering the overlap representation of the GPDs and the GTMDs in LFQM. We have listed the numerical results of $C_{z}^{q}$ for both the spin-0 and spin-1 mesons. 
There is a small variation in our results on comparing  them with the SOCs of pion and kaon in the LFQM \cite{Tan:2021osk, Kaur:2019jow}. This variation is due to the different HO parameters and normalization constants. The negative sign in the case of spin-0 mesons indicates that the OAM and quark spin tend to be anti-correlated. The positive sign in the case of spin-1 mesons  indicates that OAM and quark spin tend to be directly correlated. We presented the dependence of longitudinal momentum fraction $x$ and the transverse momentum $\mathbf{k}_{\perp}$ for the longitudinal SOC where we considered $\mathbf{k}_{\perp}$-dependence for both the GPDs and GTMDs while dealing with spin-0 mesons and only the GTMDs in the context of spin-1 mesons.  Since the spin-1 meson has an added spin wave part, we found that the dependence of longitudinal momentum fraction varies from the case of the spin-0 case. We presented fresh insights into the SOC inside the spin-0 and spin-1 mesons with our analysis of the quark longitudinal spin which has not been investigated before except for the case of pions and kaons. This work helps us understand the spin structure of different mesons. 

Future experimental data from SPD at NICA collider at JINR (Dubna, Russia) \cite{Guskov:2019qqt,Kouznetsov:2017bip} will provide deep insight into the spin structure of the hadrons. The upcoming new and upgraded experiments at JLab, DESY, EIC (electron-ion collider)
\cite{AbdulKhalek:2021gbh} will in future come up as a great source of value in accessing the spin physics data which will give extensive information to probe the multidimensional structure of hadrons.

\section{Acknowledgement}
H.D. would like to thank  the Science and Engineering Research Board, Department of Science
and Technology, Government of India through the grant (Ref No.TAR/2021/000157) under TARE
scheme for financial support.

\bibliography{biblio}

\end{document}